\documentclass[11pt,fleqn]{article}
\usepackage{amsfonts}
\usepackage{amsmath}
\usepackage{latexsym}

\usepackage{amsthm}
\usepackage[bookmarks]{hyperref}
\usepackage{comment}
\usepackage{pdfpages}
\usepackage{soul} 
\usepackage{tikz}
\usetikzlibrary{through}

\def\nottoobig#1{{\hbox{$\left#1\vcenter to1.111\ht\strutbox{}\right.\n@space$}}}
\usepackage{fancybox}
\usepackage{color}
\usepackage{xcolor}
\newcommand*{\mybox}[2]{\colorbox{#1!30}{\parbox{.95\linewidth}{#2}}}

\usepackage{times,tcolorbox}
\newtcbox{\myovalbox}[1]{colback=#1!30,boxrule=1pt,arc=5pt,  boxsep=0pt,left=3pt,right=3pt,top=3pt,bottom=3pt}

\usepackage{tikz}
\usetikzlibrary{intersections,positioning,shapes,calc,arrows,decorations.pathmorphing,decorations.pathreplacing}

\tikzset{iNode/.style={draw=blue, rectangle}}
\tikzset{fNode/.style={draw=green, circle}}
\tikzset{rNode/.style={draw=red, circle}}
\tikzset{nNode/.style={draw, circle}}

\usepackage{tablefootnote}

\newtheorem{theorem}{Theorem}[section]

\newtheorem{lemma}[theorem]{Lemma}
\newtheorem{claim}[theorem]{Claim}

\newtheorem{definition}[theorem]{Definition}
\theoremstyle{remark}
\newtheorem{remark}[theorem]{Remark}
\newtheorem*{remark*}{Remark}

\newtheorem*{question*}{Question}

\usepackage{epsfig}
\usepackage{amsmath}

\newcommand{\zo}{\{0,1\}}

\newcommand{\mapping}{\rightarrow}
\newcommand{\nat}{\mathbb{N}}
\newcommand{\F}{\mathbb{F}}

\newcommand{\cut}{\mathcal{E}}
\newcommand{\N}{{\mathcal N}}
\newcommand{\E}{{\mathcal E}}
\newcommand{\on}{\textrm{on}}
\newcommand{\off}{\textrm{off}}

\usepackage{color,soul}
\definecolor{lightblue}{rgb}{.60,.60,1}
\definecolor{lightred}{rgb}{1, .60, 0.60}
\soulregister\ref7

\newcommand{\exc}{\textrm{excess}}
\newcommand{\poly}{\textrm{poly}}

\newcommand{\ff}{{\mathrm F}}
\newcommand{\mcX}{\mathcal{X}}
\newcommand{\mcY}{\mathcal{Y}}

\if01
\fi

\makeatletter
\def\@listI{\leftmargin\leftmargini \parsep 4.5pt plus 1pt minus 1pt\topsep6pt plus 2pt minus 2pt \itemsep  2pt plus 2pt minus 1pt}

\let\@listi\@listI
\@listi
\makeatother

\author{
{Marius Zimand\/}
\thanks{  Department of Computer and Information Sciences, Towson University,
Baltimore, MD. http://orion.towson.edu/\~{ }mzimand ; The author  has been supported in part by the National Science Foundation through grant CCF 1811729.}}

\title{Online matching in lossless expanders}

\date{}
\begin{document}

\maketitle
\begin{abstract}
Bauwens and Zimand~\cite{bau-zim:t:univcompression} have shown that lossless expanders have an interesting online matching property. The result appears in an implicit form in~\cite{bau-zim:t:univcompression}. We present an explicit version of this property which is directly amenable to typical applications, prove it in a  self-contained manner that clarifies the role of some parameters, and give two applications. 

A $(K, \epsilon)$ lossless expander  is a bipartite graph such that any subset  $S$ of size at most $K$ of nodes on the left side of the bipartition has at least $(1-\epsilon) D |S|$ neighbors, where $D$ is the left degree.The  main result is that any such graph, after a slight modification, admits $(1-O(\epsilon)D, 1)$ online matching up to size $K$.  This means that for any sequence $S=(x_1, \ldots, x_K)$ of  nodes on the left side of the bipartition, one can assign in an online manner to each node $x_i$ in $S$ a set $A_i$ consisting of  $(1-O(\epsilon))$ fraction of its neighbors so that the sets $A_1, \ldots, A_K$ are pairwise disjoint. ``Online manner" refers to the fact  that, for every $i$, the set of nodes assigned to $x_i$ only depends on the nodes assigned to $x_1, \ldots, x_{i-1}$. 

The first application  concerns storage schemes for representing a set $S$, so that a membership query ``Is $x \in S$?" can be answered probabilistically by reading a single bit. Buhrman, Miltersen, Radhakrishnan and Venkatesh~\cite{bu-mi-ra-ve:c:bitvector} have shown how to design such schemes based on lossless expanders, and, subsequently, other authors have followed the same approach. All the previous one-probe storage schemes were for a static set $S$. We show that a lossless expander can be used to construct a one-probe storage scheme for dynamic sets, i.e., sets in which elements can be inserted and deleted without affecting the representation of other elements. Moreover, our method works with any lossless expander, while previous constructions required a lossless expander with a special efficient list-decoding procedure. The second application is about non-blocking networks. This is a graph that  contains  nonterminal nodes and $2N$ terminal nodes, with $N$ of them designated as input terminals, and the other $N$ designated as output terminals. The objective is, for  any $K$ pairs  of the form (input terminal, output terminal) defining a $1$-to-$1$ relation, to have  $K$ vertex-disjoint paths from the input terminal to the output terminal in each pair. Moreover, the $K$ pairs  arrive sequentially one at a time, and the paths have to be found in an online manner. $K$ is a parameter called bandwidth, and another relevant parameter is the depth which is the length of the longest path from an input terminal to an output terminal.  Using a lossless expander, we construct a non-blocking network with constant degree and almost quasilinear number of edges for a certain range of the bandwidth, improving previous constructions that had $N^{1+\Omega(1)}$ edges.

\end{abstract}

\newpage

\section{Introduction}

Expander graphs  are sparse and yet highly connected. These two apparently conflicting properties make them very useful. Avi Wigderson, in his book that comprehensively overviews  theoretical computer science, says that
expanders ``play key roles in almost every area of theory of computation: algorithms, data structures, circuit complexity, de-randomization, error-correcting codes, network design , and more.   ... In mathematics, they touch in fundamental ways different subareas in analysis, geometry, topology, algebra, number theory, and of course graph theory. ... Precious few nontrivial mathematical objects can boast a similar impact!" ~\cite[page 116]{wig:b:mathcomputation}. This paper reveals  an online matching   property of an important type of expanders, called lossless expanders, which enhances in a significant way two of the applications mentioned above (and we hope that the list will grow).

A $(K, \epsilon)$ lossless expander   is a bipartite graph $G = (L \cup R, \E)$ such that any subset  $S$ of size at most $K$ of nodes on the left side of the bipartition has at least $(1-\epsilon) D |S|$ neighbors, where $D$ is the left degree.
Repeated applications of Hall's Marriage Theorem  show that it is possible,  for every $S$ as above,  to assign to each node in $S$ a set containing an approximately $(1-\epsilon)$ fraction of its neighbors so that the sets assigned  to different nodes in $S$  are pairwise disjoint (see Section~\ref{s:expmatching}). The newly discovered property is that essentially the same can be achieved with assignments made by online matching.
The assignment procedure implied by Hall's Theorem needs to have the entire $S$.
In contrast, in \emph{online matching}, the elements of $S$ appear sequentially one at a time, and the sets that are assigned  have to be selected  before seeing future arrivals.  In other words, for a sequence  $S= (x_1, \ldots, x_K)$,  the set $A_{x_i}$ assigned to $x_i$  only depends on $\{x_1, x_2, \ldots, x_i\}$ (see Definition~\ref{d:onlinematching} for a rigorous formulation).

\begin{theorem}[{\color{blue}{Online matching in lossless expanders - informal statement. Implicit in~\cite{bau-zim:t:univcompression} }}]\label{t:main}  Let $G = (L \cup R, \E)$ be a $(K, \epsilon)$ lossless expander.
\begin{enumerate} 
\item[(a)] For every sequence $S$ of left nodes, having length $K$, there are sets $\{A_x\}_{x \in S}$ assigned online as explained above, such that each $A_x$  contains at least a fraction  $(1-O(\epsilon))$ of the neighbors of $x$, and every node in $R$ belongs to at most $O(\log |S|)$ sets.

\item[(b)]  One can slightly modify $G$ into another graph $G'$, such that for every sequence $S$ of left nodes of $G'$, of length $K$, there are \emph{pairwise disjoint} sets $\{A_x\}_{x \in S}$ assigned online as explained above, such that each  $A_x$ contains at least a fraction of $(1-O(\epsilon))$ neighbors of $x$ in $G'$.
\end{enumerate}
\end{theorem}

If $G$ is explicit,  the runtime of the online assignment procedure that assigns $A_x$ to $x$ is $\poly(K,D, \log |L|)$ in (a), and $\poly(K, D, \log |L|, 1/\epsilon)$ in (b). 

Thus, part (a) says that in a lossless expander, for every set $S$ of K left nodes,  it is possible to assign in an online manner to each node in $S$  an $(1-O(\epsilon))$ fraction of its neighbors, so that an assigned element is shared with only $O(\log |S|)$ other elements in $S$, and part (b) says, that with a slight modification of the graph,  no assigned element is shared.
\smallskip

 The online matching properties of lossless expanders   have been  observed only very recently  by Bruno Bauwens and the author~\cite{bau-zim:t:univcompression}. In that paper, the online matching algorithm  is tailored for the objectives therein, and the property appears in an implicit way. We present here a simplified and more natural  version of the online matching algorithm, with  a self-contained analysis.\footnote{\cite{bau-zim:t:univcompression} introduces the concept of an \emph{online invertible function} (see~\cite[Def. 2.1]{bau-zim:t:univcompression}), which for some settings,  is equivalent to a bipartite graph that has the online matching property. 
Theorem~\ref{t:main}(a) follows from Corollary 2.11 in ~\cite{bau-zim:t:univcompression}, and Theorem~\ref{t:main}(b) follows from  Corollary 2.13 in~\cite{bau-zim:t:univcompression} (see Appendix~\ref{s:bz19}). The procedure in~\cite{bau-zim:t:univcompression} that does the matching  is viewed from the perspective of a right node that seeks to be assigned to a left node. In this paper we use the reverse perspective and simplify the conditions of online matching and this allows us to give a version of the algorithm that is more natural for many applications of matching (like the ones we present in this paper) and that has arguably a simpler analysis.  Online matching is used  in ~\cite{bau-zim:t:univcompression} to  efficiently compress (both in the centralized and the distributed scenarios) finite strings down to almost their minimum description length.}
 Lossless expanders are closely related to lossless condensers~\cite[Th. 8.1]{ats-uma-zuc:j:expanders},  which have been studied  in the theory of pseudorandomness. We investigate online matching for general condensers, and the  results in Theorem~\ref{t:main} are obtained by particularization to the case of lossless condensers.  
 The main ideas  are  the same as in~\cite{bau-zim:t:univcompression}.

The original contributions of the paper are  two  applications that push  boundaries in the study of two basic and well-investigated problems.  
The constructions and the proofs are intuitive and simple, the reason for this being that most of the job is done by online matching.
\smallskip

The first application is about \emph{one-probe schemes} for the dictionary data structure.
The goal is to  store a  subset $S$ of a large set $U$ (the ``universe"). Let $N$ denote the size of $U$, and $K$ denote the size of $S$.  
A simple storage scheme is to keep in a table  a sorted list of the $K$ elements of $S$.  The table is stored on  $K \log N$ bits,  and, for $x \in U$, one can determine if $x$ is in $S$ or not, by reading $\lceil \log K\rceil \cdot \lceil\log N\rceil$  bits from the table.  An alternative is to have a table of $N$ bits indexed by the elements in $U$ and to set a bit to $1$ if and only its index is in $S$. Now the query ``Is $x \in S$?" can be answered by reading a single bit. Also,  one can insert or delete an element by modifying a single bit.  The cost is that the table is long (taking into account that typically $N \gg K$). 
A \emph{one-probe storage scheme} is a data structure  that answers any membership query ``Is $x$ in $S$?" by reading a single bit. 
Buhrman, Miltersen, Radhakrishnan, and Venkatesh~\cite{bu-mi-ra-ve:c:bitvector} have used  lossless expanders to build randomized one-probe storage schemes. They give both non-explicit and explicit constructions and the size of their non-explicit data structure is $ O((1/\epsilon^2) K \log N)$, where $\epsilon$ is the error probability.  Note that $K \log N$ is essentially the information-theoretical lower bound for storing the set even without the one-probe restriction. Ta-Shma~\cite{ta-shma:j:storage} and Guruswami, Umans, and Vadhan~\cite{guv:j:extractor} have obtained improved explicit one-probe storage schemes (see Section~\ref{s:storage} for parameters and other details). These one-probe storage schemes work for \emph{static} sets, in the sense that any updating of  $S$ requires the recomputation from scratch of the entire  data structure. Using the online matching property of lossless expanders, we show that each such expander yields a one-probe storage scheme for \emph{dynamic} sets.  This means that, when an element $x$ is inserted or deleted, only the bits assigned to $x$ need to be changed in the data structure, and membership queries for other elements can be answered without locking the data structure during the update. The size of the data structure depends on $K$ as before, where $K$ is now the total number of elements inserted in the dynamic set during its entire history. Plugging a condenser from~\cite{bau-zim:t:univcompression} in our construction, yields an explicit  one-probe storage scheme for dynamic sets with size $K 2^{O(\log \log (N/\epsilon) \cdot \log \log K)}$. For most $K$, this is better than the previous explicit schemes, in spite of the fact that those schemes were only handling static sets.
 The proof  is simpler and, in particular, it is noteworthy that the previous constructions of explicit one-probe storage schemes required a  lossless expander
with a special ``list-decoding" property (see~\cite[Th.7.2]{guv:j:extractor}), while our approach works with any lossless expander. Consequently, any future improvement in the construction of explicit lossless expanders will directly induce better one-probe storage schemes.
\smallskip

The second application  is about designing \emph{non-blocking networks}, which,  interestingly, has been the motivation for introducing bipartite expanders fifty years ago~\cite{bas-pin:j:networks}. In the general setting of the problem, the network has $2N$ terminals (plus non-terminals nodes as well), with $N$ of them designated as input terminals, and the other $N$ designated as output terminals, and $K$ is a parameter called \emph{bandwidth}. The objective is, for any $K$ pairs of the form (input terminal, output terminal) defining a $1$-to-$1$ relation, to have  $K$ vertex-disjoint paths from the input terminal to the output terminal in each pair.  If the $K$ pairs are known ahead of time, a graph satisfying the above requirement  is called a \emph{rearrangeable network}, and if the $K$ pairs arrive sequentially and the paths can be found in an online manner, then the graph is called a \emph{non-blocking network.}\footnote{In the literature, these networks are actually called \emph{wide-sense nonblocking networks} to distinguish them from \emph{strictly non-blocking networks}, which satisfy a stronger requirement.} The trivial non-blocking network is the complete bipartite graph with $N$ nodes on each side, which has depth $1$,  but $N^2$ edges (the depth is the length of the longest path from an input terminal to an output terminal).  In our application, we show that any $(K, (1-\epsilon)D)$ expander $G = (L \cup R, \E)$ with $|L| = N$ and left degree $D$  yields via a very simple construction a non-blocking network with $ND + |R|^2$ edges, and depth $3$. If we use a condenser from~\cite{bau-zim:t:univcompression} and if $K = O(N^{1/2})$ we obtain a non-blocking network with an almost quasi-linear number of edges, while, as far as we know, previous methods are only able to obtain $N^{1 + \Omega(1)}$ many edges  for constant-depth non-blocking networks.  The online matching of lossless expanders is a quite general tool and by mixing it with known constructions one can obtain non-blocking networks with other interesting combinations of parameters.  Our goal, however, is to just illustrate the method, and we do not pursue here this line of investigation.

\subsection{Basic definitions and notation.} \label{s:defs}
We restrict our attention  to bipartite graphs $G= (L \cup R, \E)$ with left degree $D$, i.e, the nodes are partitioned into the set of left nodes $L$ and the set of right nodes $R$, all edges connect a left node with a right node, and all left nodes are adjacent to exactly $D$ edges. We allow $\E$ to be a multiset (i.e., it is possible to have several edges between two vertices).\footnote{We use calligraphic fonts such as $\E, \N$ to denote multisets.} We label the edges adjacent to a left node $x$ by a value in $[D] = \{1,2, \ldots, D\}$, and sometimes we view the graph as a function $\Gamma : L \times [D] \mapping R$ defined by $\Gamma(x, y) = p$ if $(x,p)$ is an edge in $G$ labeled with $y$.
For every node $x$, we denote by $\N(x)$ the \emph{multiset} of neighbors of $x$. If there are $\ell$ edges $(x,p)$, then $p$ has multiplicity $\ell$ in $\N(x)$. Thus, for every $x \in L$, the size of $\N(x)$ is $D$. For $S \subseteq L$, we define the neighborhood set  $N(S) := \{v \mid \exists u \in S, (u,v) \in E\}$. For $A \subseteq L, B \subseteq R$, $\cut(A,B)$ denotes the multiset of edges $(u,v)$ with $u \in A$ and  $v \in B$, i.e.,  the multiset of edges that cross from $A$ to $B$. 

We use \emph{bipartite vertex expanders} (see~\cite[Def. 4.3]{vad:b:pseudorand}), which, henceforth, we simply call expanders.
\begin{definition}[{\color{blue}{ expander }}] \label{d:expander}
A graph as above is a $(K, \gamma)$ expander if for every set $S \subseteq L$ of size at most $K$, $|N(S) | \ge \gamma |S|$.
\end{definition}
Thus $(K,  (1-\epsilon)D)$ expander  is the same notion as $(K, \epsilon)$ lossless expander, introduced earlier.

The number of occurrences of an element in a multiset $A$ is called the \emph{multiplicity} of the element. The size of $A$, denoted $|A|$, is the sum of multiplicities. For example,  $|\{a,a,b\}| = 3$. For multisets $A, B$, $A \subseteq B$ means that the multiplicity of any element in $A$ is at most its multiplicity in $B$. ${\cal P}(A)$ is the powerset of $A$ and ${\cal P}_{\textrm{multi}}(A)$ is the set of multisets with elements from $A$.
\medskip

\subsection{Offline vs. online matching} \label{s:expmatching}
It is useful to have a general version of matching for bipartite graphs, that we dub $(\ell, r)$ matching.   Such a matching assigns to every left node at least $\ell$ of its neighbors, so that  every right node is assigned  to at most $r$  left nodes.  The larger is $\ell$ and the smaller is $r$, the stronger is the property of  having an $(\ell, r)$ matching. 

Let us see an example. 
Let $G$ be a $(K, \gamma D)$ expander. Assume that $\gamma D \geq 1$. 
 Let $S$ be a subset of left nodes of size at most $K$. Since $|N(S')| \geq |S'|$ for every $S' \subseteq S$, by Hall's  Marriage Theorem,  the graph obtained from $G$ by restricting the left side to $S$ has an exact matching (i.e., there is a subset of $|S|$ edges, defining a 1-to-1 relation). We assign to each node in $S$,  the right node with whom it is matched.  By repeating this process $\lfloor \gamma D \rfloor$ rounds
(where after each round we remove the right nodes that have been matched), we can assign to each node in $S$,  $\lfloor \gamma D \rfloor$ of its neighbors, so that every  right node is assigned to at most one node in $S$.  In other words,  $G$  restricted to $S$ has  $(\gamma D, 1)$  matching, for every subset $S$ of $K$ left nodes..
In particular, a $(K, \epsilon)$ lossless expander is by definition a $(K, (1-\epsilon)D)$ expander and thus it has the very strong property of $(\lfloor (1-\epsilon) D\rfloor, 1)$ matching when we restrict the left side to $S$ of size $K$, for any such $S$.  Our main result is that, essentially, the same is true for online matching.

We next  define online matching in bipartite graphs.
We start with an informal discussion, which is meant to help the interpretation of the formal definition.   In online matching, a left node may make a request to get assigned to it many of its neighbors, so that an assigned neighbor is  shared with only few other nodes, in the sense of $(\ell, r)$ matching.  The assignment requests arrive and our definition also allows that they depart (but in a restricted  way), and each request must be  satisfied when it arrives before seeing future arrivals. The set assigned to a node must not change between the arrival and the departure time of the node. The requests are specified by a list $S$ of left nodes, i.e., a sequence $S= (x_1, \ldots, x_K)$ with every $x_i \in L$, which can be updated as a \emph{stack}.   A ``request arrival" means that an element  is pushed in $S$. A   ``request departure" means that an element is popped from $S$, in the stack manner. Thus, we assume that the top of the stack $S$ is the last position, and, when a new element $x$ is inserted,  $S$ becomes $(x_1, \ldots, x_K, x)$, and  only the last element can be deleted.  

 The formal definition is as follows.

\begin{definition}[{\color{blue}{$(\ell,r)$ online matching}}  ] \label{d:onlinematching}
Let $G = (L \cup R, \E)$ be a bipartite graph, $K \in \nat$, and  let ${\cal L}^{\le K}$ denote the set of lists of elements in $L$ of size at most $K$. The graph $G$ admits  $(\ell, r)$ online matching   up to size $K$ if there is a function $f : {\cal L}^{\le K} \times L \mapping {\cal P}_{\textrm{multi}}(R)$  such that for every $S\in {\cal L}^{\le K}$  and for every $x \in L$,
\begin{enumerate}
\item $f(S, x) \subseteq \N(x)$,
\item If $x \not\in S$, then $f(S,x) = \emptyset$,
\item If $x \in S$, then for every list $S' \in {\cal L}^{\le K}$ that extends $S$, $f(S, x) = f(S', x)$, 
\item If $x \in S$, then $|f(S, x)| \ge \ell$,  and
\item For every $p \in R$, $|\{x \in S \mid p \in f(S,x) \}| \le r$.

\end{enumerate} 
\end{definition}
  Lists $S \in {\cal L}^{\le K}$ are interpreted as snapshots of the stack of requests at various moments.  
The assignment function $f$ assigns to every  left node in $S$  at least $\ell$ of its neighbors (using multiplicities in the count), and every right node is assigned to at most $r$ different neighbors.The online matching is stipulated in property (3), which  implies  that the set of elements assigned to $x$ does not change for the entire lifetime of $x$ in  the stack of requests. 
\medskip

Online matching is a stronger requirement than offline. For example, the following graph admits $(1,1)$ offline matching up to size $K=2$, but not $(1,1)$ online matching up to size $K=2$ (consider the case when $x_2$ arrives first, and the second arrival is chosen adversarially after a  node was assigned to $x_2$).

\medskip
  
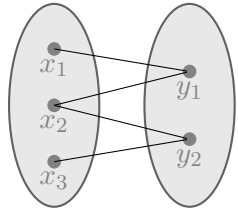
\begin{center}
\begin{tikzpicture}[shorten >=1pt,scale=0.15,
lnode/.style={fill=black, circle, scale=0.5} 
]

\coordinate (x) at (0,0);
\coordinate (y) at (12,0);
\coordinate (l1) at (0,-5);
\coordinate (l2) at (0,0);
\coordinate (l3) at (0,5);
\coordinate (r1) at (12,-3);

\coordinate (r2) at (12,3);

\node[lnode] (c3) at  (l3){};
\node[lnode] (c2) at  (l2){};
\node[lnode] (c1) at  (l1){};
\node[lnode] (d1) at  (r1){};
\node[lnode] (d2) at  (r2){};

\node[anchor=north] (cc3) at (l1.-90) {$x_3$};
\node[anchor=north] (cc2) at (l2.-90) {$x_2$};
\node[anchor=north] (cc1) at (l3.-90) {$x_1$};

\node[anchor=north] (cd2) at (r1.-90) {$y_2$};
\node[anchor=north] (cd1) at (r2.-90) {$y_1$};

\filldraw[opacity=0.6, color=black, fill = gray!30, thick ] (x) ellipse (4 and 9);
\filldraw[opacity=0.6, color=black, fill = gray!30, thick ] (y) ellipse (4 and 9);
\draw[-] (l1) -- (r1);
\draw[-] (l2) -- (r1);
\draw[-] (l2)--(r2);
\draw[-] (l3)--(r2);

\end{tikzpicture}
\end{center}

\subsection{From online matching with sharing to online matching with no sharing}\label{s:nosharing}

We convert any graph $G$  that admits $(\ell, r)$ online  matching to a graph $G'$  that admits  $(\ell', 1)$ online matching such that $\ell/D \approx \ell'/D'$  ($D$ and $D'$ are the left degrees of $G$, respectively $G'$) and
 without affecting too much the left degree and the size of the right side. The idea is to use hashing  to distinguish between the left nodes that share a right node. 
\smallskip

\textbf{The $G \mapsto G'$ transformation.}
\smallskip
 
\noindent
 Let $G = (L \cup R, \E)$ be a graph with left degree $D$ that admits $(\ell, r)$ online matching up to size $K$.  Let $n = \lceil \log |L| \rceil$ and let $t$ be the smallest power of two that is at least $(1/\epsilon) (n-1)(r-1)$, for some parameter $\epsilon > 0$. We use hashing via polynomials of low degree and for this we label all the  left nodes  by vectors in $\F_{2^n}$.  We view each $x \in L$ as a polynomial of degree at most $n-1$ over $\F_t$ in the natural way, by considering each bit in the label of $x$ as a coefficient of the polynomial (the polynomial $x$ has only $0,1$ coefficients; $\F_2$ and $\F_t$ are the finite fields with $2$, respectively $t$ elements).

We construct the bipartite graph $G' = (L' \cup R', \E')$. The left side $L'$  is $L$ (so, the same left side as $G$). The right side is $R' = R \times (\F_t)^2$. The multiset of  edges $\E'$  of $G'$ is defined as follows:

 For each edge $(x, p)$ of $G$ (where $x$ is a left node, and $p$ is a right node in $G$), we introduce $t$ edges in $G'$, namely $\{(x,\, (p, x(a), a) )\mid  a \in \F_t\}$, where $x(a)$ is the value of the polynomial $x$ at $a$.

Note that $t = O((1/\epsilon) r \log |L|)$,   the graph $G'$ has left degree $D' = D \cdot t$, and the size of the right side is $|R'| = |R| \cdot t^2$. Also note that if $G$ is explicit, $G'$ is explicit as well.\footnote{We remind that $G$ is explicit if it  belongs to a family of graphs indexed by $\log |L|$ and there exists  an algorithm running in time $\poly(\log |L|)$ that, on input $x \in L$ and  $i \in [D]$, outputs the $i$-th neighbor of $x$.}
\begin{lemma}[{\color{blue}{$G \mapsto G'$ transformation}}]\label{l:noshare}
If $G$ admits $(\ell, r)$ online matching up to size $K$, then $G'$ admits $((1-\epsilon) \ell t, 1)$ online matching up to size $K$. 
\end{lemma}
\begin{remark} \label{r:noshare}
In particular, if   $G$ admits $((1-\epsilon') D, r)$ online matching, then $G'$ admits $((1-\epsilon - \epsilon') D', 1)$ online matching.
\end{remark}
\begin{proof}
We modify the assignment procedure $f(S,x)$ for $G$ into an assignment procedure $f'(S, x)$ for $G'$. Let $S = (x_1, \ldots, x_K)$ be a list of left nodes  (recall that the order in the list is interpreted as the  order of ``arrival"). 
\smallskip

The assignment procedure $f'(S,x)$ works as follows. 
\smallskip

First we execute $f(S,x)$. Each right node $p$ that $f$ assigns to $x$ (i.e., $p \in f(S,x)$) may also have been assigned to other $r' \leq r-1$   elements that have arrived in $S$ before $x$, say,   to $\{x_1, \ldots, x_{r'}\} \in S - \{x\}$. For every $i \le r'$, the polynomials $x$ and $x_i$ can be equal in at most $(n-1)$ points in $\F_t$, because they have degree at most $n-1$. Thus there exists a set $A$ of at least $t - (n-1)(r-1) \geq (1-\epsilon)t$ points in $\F_t$ such that for every $a \in A$,  $x(a) \not\in \{x_1(a), \ldots, x_{r'}(a)\}$.   Then $f'$ assigns to $x$ the elements $\{(p, x(a), a) \mid a \in A\}$ (for all $p \in f(S,x)$). By the above estimations,  $f'$ assigns to each left $x$ at least $(1-\epsilon) \ell t$  of its neighbors, and the assignment procedure ensures that no right node is assigned to more than one left node. 
\end{proof}

\begin{remark}\label{r:timenoshare}The assignment procedure for $G'$ runs the assignment procedure for $G$ and next calculates  $x(a)$ for all $x \in S$ and all $a \in \F_t$. The evaluations take time $\poly(K, n,  t)  = \poly(K, n, r, 1/\epsilon)$ (because $t = O(1/\epsilon \cdot n \cdot r)$). Therefore the running time of the assignment  procedure for $G' =$ running time of the assignment  procedure for $G$ $+ \poly(K,n,r, 1/\epsilon)$. 
\end{remark}

 \section{Online matching in lossless expanders and related graphs}\label{s:onlinematching}


It would be interesting to find a property of bipartite graphs that  characterizes graphs admitting online matching, similarly to Hall's Marriage Theorem for offline matching. We do not solve this problem, but we do identify a property that is sufficient for a strong type of online matching.

  Clearly, if in a bipartite graph $G$ with left degree $D$,  for every subset $S$ of left nodes, of size at most $K$, every node is adjacent to at most $r$ edges coming from  $S$, then $G$ admits $(D, r)$ online matching up to size $K$ (we simply assign to each $x$ in $S$, all its neighbors). Unfortunately, only graphs $G$ with   large right side $R$ can have this property.

We introduce a relaxed version of the above  property, which we dub \emph{$( r, K,  \epsilon)$ bounded right degree}.  Informally, the property requires that for every subset $S$ of left nodes,  of size at most $K$, if we discard $\epsilon D |S|$ edges, then every right node is adjacent to at most  $r$ edges coming from  $S$.  

Formally, for each subset $S$ of left nodes, every natural number $r$, and for every right node $p$, $|\cut(S,p)|$ denotes the size of the multiset of edges  crossing from $S$ to $p$,  and we define 
\[
\begin{array}{ll}
\exc_S(p,r) & = \max ( |\cut(S, p)| - r, 0) \\
\\

\exc_S(r) & = \sum_{p \in N(S)} \exc_S(p, r).
\end{array}
\]

\begin{definition}[{\color{blue}{$ (r, K,  \epsilon)$ bounded right degree}}  ] \label{d:bounddeg}
A bipartite graph $G = (L \cup R, \E)$ with left degree $D$  has \emph{$(r,  K, \epsilon)$ bounded right degree}  if for every $S \subseteq L$ of size at most $K$,
$\exc_S(r) \le \epsilon D |S|$.
\end{definition}
\begin{remark}\label{r:expboundeddeg}
It is easy to check that $G$ has $(1, K, \epsilon)$ bounded right degree if and only if $G$ is a $((1-\epsilon)\cdot D, K)$-expander.
\end{remark}

\noindent
We make two claims: The property of $(r, K, \epsilon)$ bounded right degree,  
\medskip

\noindent
{\color{black}{( *)}} is sufficient for $((1-O(\epsilon))D, O(\log K \cdot r))$ online matching up to size $K$, and   
\medskip

\noindent
{\color{black}{(**)}} characterizes \emph{condenser graphs}, a type of graph that has been studied in the theory of pseudorandomness and that can be viewed as a generalization of lossless expander graphs.  
\medskip

\noindent
Claim (**)  is useful because there are constructions of explicit condenser graphs in which the size of $R$ is not much larger than $K$.
We prove the two claims above and after that the main results of this section.
\smallskip

\subsection{Proof of claim (*)}

In this section, $G = (L \cup R, \E)$ is  a bipartite graph with left degree $D$ that  has $(r, K,  \epsilon)$ bounded right degree.
We need two  concepts.
\begin{definition}[{\color{blue}{ heavy / deficient nodes}}  ] \label{d:heavydef}
 Let $S$ be a subset of left nodes.
\begin{itemize}
  \item A right node $p$ is \emph{heavy for $S$} if it has more than $2r$ different neighbors in $S$.
\item A left node $x$ in $S$ is \emph{deficient for $S$} if $| \cut(x, \textrm{HEAVY}) | \ge 4 \epsilon D$, where $\textrm{HEAVY}$ is the set of nodes that are heavy for $S$.
\end{itemize}
\end{definition} 

The following is the key property of $G$  that is used for online matching. 
\begin{lemma}\label{l:defic}
Let $S$ be a subset of left nodes, of size at most $K$. Then the subset of elements deficient for $S$ has size at most $|S|/2$.
\end{lemma}
\begin{proof} Let $\textrm{HEAVY}$ be the set of nodes that are heavy for $S$, and let $\textrm{DEFICIENT}$ be the set of nodes that are deficient for $S$.  We color the edges in $\cut(S,R)$, i.e., the edges going out from $S$. The other edges are ignored in the rest of the proof. For each right node $p$,  we color in red 
$\exc_S(p, r)$ edges adjacent to $p$  and color in green the other edges. In other words, the red edges are the edges that we ``discard" so that each right node remains with  at most $r$ green edges coming from $S$. Since the number of red edges is at most $\epsilon D |S|$ and for each  heavy $p$ we color in red more than $r$ adjacent edges, it follows that 
\[
|\textrm{HEAVY}| <  (1/r) \cdot \epsilon  D |S|.
\] 

Suppose $|\textrm{DEFICIENT}| > |S|/2$. Then  the total number of edges that cross from $\textrm{DEFICIENT}$ to $\textrm{HEAVY}$ is greater than $|S|/2 \cdot 4 \epsilon D = 2 \epsilon D |S|$.  It follows that the number of green edges  adjacent to \textrm{HEAVY} is greater than $\epsilon D |S|$  (because there are at most $\epsilon D |S|$ red edges and the rest are green).  Since each node has at most $r$ green edges adjacent to it, it follows that 
\[
|\textrm{HEAVY}| > (1/r) \cdot \epsilon D |S|.
\]
 This contradicts the previous inequality, and ends the proof.
\end{proof}

We are ready to prove claim {\color{blue}{(*)}}.
\begin{theorem}[{\color{blue}{ bounded right degree $\Rightarrow$ online matching}}  ] \label{t:onlinematching}
If $G = (L \cup R, \E)$ is a bipartite graph with left degree $D$ that  has $(r, K,  \epsilon)$ bounded right degree, then $G$ admits $((1-4\epsilon) D, 2 \lceil \log K \rceil r)$ online matching up to size $K$. 

Furthermore, if the graph $G$ is explicit, then there is an algorithm for the assignment function $f(S,x)$ with running time $\poly(K, D, \log |L|)$.
\end{theorem}
\begin{proof}
Let us consider a list of left nodes $S$ of size at most $K$.  If $x$ is not in $S$, we define $f(S,x) = \emptyset$. For $x \in S$, we define $f(S,x)$ by the following procedure.
\medskip

  \newcommand{\alg}{\quad\quad  \textbf{Computation of $f(S,x)$ }
\medskip

  $j \leftarrow $ the ordinal of the first occurrence of $x$ in the list $S$;  $S_0 \leftarrow \textrm{set of the first $j$ elements in  $S$}$;  $t \leftarrow 0$; 

while ($x$ is deficient for $S_t$) 

\quad\quad\quad $S_{t+1} \leftarrow$ the set of elements in $S_t$ that are deficient for $S_t$;

\quad\quad\quad $t \leftarrow t+1$

end-while

$f(S, x) \leftarrow$ the multiset of non-heavy for $S_t$ neighbors of $x$ (with their multiplicity from $\N(x)$)

}
\mybox{gray}{\alg}
\medskip

We check that $f$ defined in the above procedure satisfies the requirements (1)-(5) in the Definition~\ref{d:onlinematching}. The first two follow immediately. 

Requirement (3) is satisfied because 
the computations of  $f(S, x)$ and $f(S', x)$  start with the same $S_0$ and are therefore identical.

We move to (4). The loop terminates in at most $\lceil \log K \rceil$ iterations, because, by Lemma~\ref{l:defic}, $|S_{t+1}| \le |S_t|/2$. If $S_t$ decreases to just $2r$ elements, then those elements cannot be deficient (because there cannot exist heavy nodes for such $S_t$). Thus,  $x$ eventually becomes non-deficient, and when this happens, it has at least $(1-4\epsilon)D$ non-heavy neighbors (including the multiplicity in the count). Thus $f$ assigns to $x$ a multiset with $(1-4\epsilon) D$ of its neighbors.

 It remains to check (5), i.e., to show that every right node is assigned to at most   $2 \lceil \log K \rceil r$ nodes in $S$. Let $p$ be a right node and $S = (x_1, x_2, \dots, x_K)$.   We look at all computations $f(S, x_1), \ldots, f(S, x_K)$ and estimate in how many of them $p$ is assigned. 
\begin{claim}\label{c:bound}
 $p$ is  assigned to at most  $2r$ elements at any given iteration of the computations of   $f(S, x_1), \ldots, f(S,x_K)$.
\end{claim}
 \begin{proof}
We analyze an arbitrary  iteration $t^{*}$ of the assignment procedures for all   $x_1, \ldots, x_K$.  Suppose that $p$ is assigned at iteration $t^{*}$  in  $2r+1$ of these procedures, to elements $x_{i_1}, \ldots, x_{i_{2r+1}}$, where $i_1 < \ldots < i_{2r+1}$. Note that  $t^{*}$ is the value of the parameter $t$ when the while loop terminates in all computations $f(S, x_{i_1}), \ldots, f(S,x_{i_{2r+1}})$.  It means that each $x_{i_j}$ is a neighbor of $p$ and that it was deficient at iteration $t^{*}-1$ in the computation of $f(S, x_{i_j})$.  But then every $x_{i_j}$ is also deficient at iteration $t^{*}-1$ in the computation of  $f(S, x_{i_{2r+1}})$, because the computations start with increasingly larger $S_0$, and  the predicates ``heavy for $S$" and  ``deficient for $S$" are monotonous in $S$ (once an element becomes heavy for some list  $S'$, it remains heavy for every list $S^{''}$ that extends  $S'$).  It follows that  $x_{i_1}, \ldots, x_{i_{2r+1}}$ are all in $S_{t^*}$ in the computation of $f(S, x_{i_{2r+1}})$ and are all neighbors of $p$. On the other hand,  in this last computation,  $p$ is assigned to $x_{i_{2r+1}}$ and thus  $p$ has at most $2r$ neighbors    in $S_{t^*}$, because it is not heavy. We have reached a contradiction and the claim is proved.
\end{proof}
As we have noticed earlier,  for each $x_i$ in $S$, the algorithm that computes $f(S, x_i)$ has at most  $\lceil \log K \rceil$ iterations. Therefore, using  Claim~\ref{c:bound}, we infer that every right node can be assigned to at most   $2 \lceil \log K \rceil r$ nodes in $S$.  The claimed running time  follows by a straightforward inspection of the algorithm.
\end{proof}

\subsection{Proof of claim (**)}

Condensers are a type of functions, studied in the theory of pseudo-randomness (see~\cite{vad:b:pseudorand}),  that play an important role in establishing the online matching property of lossless expanders.  We present their definition. A random variable has {\em min-entropy $k$} if each value has probability at most $2^{-k}$. The statistical distance between two random variables $P$ and $Q$ with the same range is $\sup |P(S)-Q(S)|$, with the supremum taken over all subsets $S$ of the range. For $\epsilon > 0$, we say that $P$ and $Q$ are $\epsilon$-close, if their statistical distance is bounded by $\epsilon$.
 Given a set $B$, we denote $U_B$  a random variable that is uniformly distributed on $B$. 

\begin{definition}[{\color{blue}{ condenser }}  ] \label{d:condenser}
 A function $C: \zo^n \times \zo^d \mapping \zo^{m}$ is a $k \rightarrow_{\epsilon} k'$ condenser, if for every random variable $X$ with min-entropy at least $k$ (ranging over $\zo^n$), the random variable 
$Y= C(X, U_{\zo^d})$ is $\epsilon$-close to a random variable $\widetilde{Y}$ that has min-entropy at least $k'$. 
\end{definition}

A condenser is thus a randomized transformation of random variables $X \mapsto Y$ ($U_{\zo^d}$, where $d$ is typically small,  is an auxiliary random  variable representing the randomness of the transformation).  For typical settings of parameters, a condenser enhances randomness in the sense that the output $Y$ is closer to having  uniform distribution than   the input $X$.   The quantity $ k + d - k'$ is called the \emph{entropy loss} of the condenser, because the input has min-entropy $k+d$ and the output  is close to having min-entropy $k'$.  
We view $C$ as a bipartite graph $G$ in the usual way: the left nodes are the strings in $\zo^n$, the right nodes are the strings in $\zo^m$ and for each $x \in \zo^n$, $\rho \in \zo^d$ there is an edge $(x, C(x, \rho))$.

We actually work with functions that have the condenser property for a large range of values of $k$.  Namely, we use families of functions  $C$ indexed by $n$ (but as usual we do not write the index) of the following type 
\begin{equation*}
\begin{array}{l}

\text{\color{black}{(***) }\quad\quad } \text{$C: \zo^n \times \zo^d \mapping \zo^m$ is a $k \rightarrow_\epsilon k+d-e$ condenser} \\
\text{\quad\quad\quad\quad for all $k \le k_{max}$ such that $2^k \in \nat$.}

\end{array}
\end{equation*}
Here the parameters $d, m, \epsilon,  k_{max}$, and $e$ are functions of $n$, and with the exception of $\epsilon$ are positive integers.  Functions of type~(***) are very similar to \emph{conductors}, the difference being that conductors do not have the restriction that $2^k \in \nat$. The parameter $e$ is a bound of the entropy loss for all $k \le k_{max}$ and plays an important role for the online matching property.
 
\begin{lemma}[{\color{blue}{ condenser $\Leftrightarrow$ bounded right degree }}  ] \label{l:condenserdegree}
A function $C$ has parameters as indicated in condition (***) if and only if the corresponding graph has $(2^e, 2^{k_{max}}, \epsilon)$ bounded right degree and $m \ge k_{max}+d-e$.
\end{lemma}
\begin{proof}
`` $\Rightarrow$"  Let $S \subseteq L$ with size $|S| \le 2^{k_{max}}$. Let $D = 2^d$ and $k = \log |S|$. Then $U_S$ has min-entropy $k \le k_{max}$, and $Y = C(U_S, U_{\zo^d})$ is $\epsilon$-close to a random variable $Y'$ with min-entropy bounded by $k+d - e$.  Clearly, $m \ge k+d-e$ because otherwise no random variable with  range $\zo^m$ can have min-entropy $k+d-e$. We need to show that 
$\exc_S(2^e) \le \epsilon D |S|.$

Let $A = \{ p \in N(S) \mid |\cut(S,p)| > 2^e\}$. We have
\[
\begin{array}{ll}
\epsilon & \ge Y(A) - Y'(A) = \sum_{p \in A} Y(p) - Y'(p) 
      \ge \sum_{p\in A} \frac{|\cut(S,p)|}{|S| \cdot D} - 2^{-(k+d-e)} \\ \\
& =  \frac{1}{|S| \cdot D} \sum_{p \in A} (|\cut(S,p)| - 2^{e}) =   \frac{1}{|S| \cdot D} \sum_{p \in A} \exc_S(p, 2^e) = \frac{1}{|S|D} \exc_S(2^e).
\end{array}
\]
which implies the desired inequality.
\smallskip

``$\Leftarrow$" It is well known that it is enough to show the condenser property for all flat distributions $X$ (see~\cite[Lemma 6.10]{vad:b:pseudorand}). So we take $X$ to be $U_S$, where  $S \subseteq \zo^n$ with size $|S| = 2^k$ for some $k \le k_{max}$. Let $Y = C(U_S, U_{\zo^d})$.  
\begin{claim}In the graph that corresponds to $C$, it is possible to  redirect $\epsilon D |S|$ edges so that after redirection every element in $\zo^m$ has at most $2^e$ neighbors in $S$.\footnote{Redirection means changing the right endpoint of the edge.}
\end{claim} 
\begin{proof}
Let $A$ be the set of elements $p \in \zo^m$ such that $|\cut(S,p)| > 2^e$ and let $\overline{A} = \zo^m - A$.  Let $\textit{excess}$ be the number of edges that need to be ``shaved" so that all nodes in $\zo^m$ get to have $2^e$ edges coming  from $S$. We have $\textit{excess} \le \epsilon D |S|$, because the graph has $(2^e, 2^{k_{max}}, \epsilon)$ bounded right degree.  We define \emph{deficit} symmetrically to  \emph{excess} (i.e., the \emph{deficit} is the number of edges that need to be added to get the same condition). The number of edges adjacent to $A$ is 
$|A| \cdot 2^e + \textit{excess}$ and the number of edges adjacent to $\overline{A}$ is $|\overline{A}| \cdot  2^e - \textit{deficit}$. Since the number of edges adjacent to $A \cup \overline{A}$ is $D \cdot |S|$ and $|A \cup \overline{A}| = 2^m$, it follows that $2^m \cdot 2^e + \textit{excess} - \textit{deficit} = D\cdot |S|$, which implies that $\textit{excess} \le \textit{deficit}$ (taking into account that $m \ge k_{max}+d-e$). Therefore we can redirect \emph{excess} many edges coming from $S$  so that instead of ending in $A$ they end in $\overline{A}$, and no right node has more than $2^e$ edges adjacent to it. 
\end{proof}
The redirection yields a random variable $Y'$ with range $\zo^m$, with min-entropy at least $k+d - e$, and $Y'$ is $\epsilon$-close to $Y$ because at most a fraction of $\epsilon$ edges have been redirected.
\end{proof}

\subsection{Proofs of the online matching properties of lossless expanders and condensers}

We are now prepared to present and prove the results announced in the Introduction.
\begin{theorem}[{\color{blue}{ online matching properties of  lossless expanders  (with sharing)}}  ] \label{r:lexpamatching}
If $G = (L \cup R, \E)$  is a bipartite graph with left degree $D$ that is  a $(K_{max}, (1-\epsilon) D)$ expander, then  $G$ admits $((1-4 \epsilon) D, 2 \lceil \log K_{max} \rceil  )$ \emph{online} matching up to size $K_{max}$.
\medskip

\noindent
If $G$ is explicit, then the assignment procedure $f(S,x)$ from Theorem~\ref{t:onlinematching} for $G$ has running time  \\
$\poly(K_{max}, D, \log |L|)$. 
\end{theorem}\label{t:main-no-sharing}
\begin{proof}
By Remark~\ref{r:expboundeddeg}, $G$ has $(1, K_{max}, \epsilon)$ - bounded right degree (this also follows from Ta-Shma, Umans, and Zuckerman~\cite[Th. 8.1]{ats-uma-zuc:j:expanders} where it is shown that $G$ is a $k \rightarrow_\epsilon k+d$ condenser for every $k \le \log K_{max}$ such that $2^k \in \nat$). So, Theorem~\ref{t:onlinematching} implies that $G$ admits $((1-4 \epsilon) D, 2 \lceil \log K_{max}\rceil)$ online matching up to size $K_{max}$.
\end{proof}
\begin{theorem}[{\color{blue}{ online matching properties of  lossless expanders (with no sharing)}}  ]  \label{t:main sharing}
If $G = (L \cup R, \E)$  is a bipartite graph with left degree $D$ that is  a $(K_{max}, (1-\epsilon) D)$ expander, then the transformation $G \mapsto G'$ yields
a graph  $G' = (L' \cup R', \E')$ with left degree $D'$ that admits $((1-5 \epsilon) D', 1)$ \emph{online} matching up to size $K_{max}$, and  $G'$ has parameters $L' = L$, $|R'| = |R| \cdot t^2$, $D' = D \cdot t$, for $t = O((1/\epsilon) \cdot \log |L| \cdot \log K_{max})$.
\medskip

\noindent
If $G$ is explicit, then $G'$ is explicit and the assignment procedure $f(S,x)$ from Theorem~\ref{t:onlinematching} combined with the $G \mapsto G'$ transformation for $G'$ has running time  
$\poly(K_{max}, D', \log |L|, 1/\epsilon)$. 

\end{theorem}
\begin{proof}
The statement follows by combining Theorem~\ref{t:main-no-sharing} and the properties of the $G \mapsto G'$ transformation from Section~\ref{s:nosharing}.
\end{proof}
\medskip

\begin{theorem}[{\color{blue}{ condenser $\Rightarrow$ online matching (with sharing) }}  ] \label{t:condensermatching}
If a function $C$ has parameters as in condition (***), then the corresponding graph admits $( (1-4\epsilon)2^d, 2 \cdot k_{max}   \cdot 2^e)$ online matching up to size $2^{k_{max}}$. 

\noindent
If $C$ is explicit, then the running time of the assignment procedure $f(S,x)$ from Theorem~\ref{t:onlinematching} is $\poly(2^{k_{max}}, 2^d, n)$.
\end{theorem}
\begin{proof}
This follows by combining Theorem~\ref{t:onlinematching} and Lemma~\ref{l:condenserdegree}
\end{proof}

\begin{theorem}[{\color{blue}{ condenser $\Rightarrow$ online matching (with no sharing) }}  ] \label{t:condenser-nosharing}
If a function $C$ has parameters as in condition (***) and $G = (L \cup R, \E)$ is the graph corresponding to $C$,  then the transformation $G \mapsto G'$ yields
a graph  $G' = (L' \cup R', \E')$ with left degree $D'$ that admits $((1-5 \epsilon) D', 1)$ \emph{online} matching up to size $K_{max}$, and  $G'$ has parameters $L' = \zo^n$, $|R'| = 2^m  \cdot t^2$, $D' = 2^d \cdot t$, for $t = O((1/\epsilon) \cdot \log |L| \cdot \log K_{max} \cdot 2^e)$.

\noindent
If $C$ is explicit, then $G'$ is explicit and the running time of the assignment procedure $f(S,x)$ from Theorem~\ref{t:onlinematching} is $\poly(2^{k_{max}}, 2^d, n, 1/\epsilon)$.
\end{theorem}
\begin{proof}
The statement follows by combining Theorem~\ref{t:condensermatching} and the properties of the $G \mapsto G'$ transformation from Section~\ref{s:nosharing}.
\end{proof}
\medskip

In short, expander graphs and  condenser graphs of type (***) yield bipartite graphs $G$ (or $G'$, after the simple transformation $G \mapsto G'$) with left size $L = \zo^n$, that have good online matching up to size $K_{max} := 2^{k_{max}}$. In typical applications, $n$, $k_{max}$ and $\epsilon$ are given, and it is important that the graph has a right side $R$, with size not much larger than $K_{max}$. More precisely, we refer to  $\delta = \log (|R| / K_{max})$ as \emph{overhead}, and it is desirable to have $\delta =  \poly(\log(n/\epsilon))$, or even  $\delta =  O(\log(n/\epsilon))$. It is also desirable that $d$ is small, even though this appears  to have somewhat less impact.  In most applications it is important  that the condensers are \emph{explicit}. We remind that this means that we have a family of condensers indexed by $n$, and that $C(x,y)$ for $x$ of length $n$ is computable in time polynomial in $n$.
\medskip

Table~\ref{t:tablecondenser} presents some condensers from literature, and the online matching property of the corresponding graph $G$. We also indicate the condensers that are strong, and those that are linear, which are useful properties in some applications. A  $k \rightarrow_\epsilon k'$ condenser $C : \zo^n \times \zo^d \mapping \zo^m$  is \emph{strong}  if, either  the output contains the seed at some fixed coordinates, or, if by concatenating the output with the seed,  it becomes a $k \rightarrow_\epsilon k'+d$ condenser.
A condenser is \emph{linear}, if for every fixed seed $y \in \zo^d$, the function $C(\cdot, y)$ is linear, i.e., $C(x+x', y) = C(x,y) + C(x',y)$, for all $x,x'$ (`+` is bitwise XOR; in other words, $n$-bit strings are viewed in the natural way as elements of the additive group $\F_2^n$; for short proof sketches of the linearity of these condensers,  see for example the appendix of~\cite{bau-zim:t:univcodes}).
\medskip

Table~\ref{t:exponlinematching} presents the corresponding bipartite graphs $G'$  obtained via the transformation $G \mapsto G'$ from Section~\ref{s:nosharing}, which admit online matching with no sharing. If we consider the functional view of a bipartite graph, the expanders (2), (3) and (5) are linear, because the transformation $G \mapsto G'$ preserves linearity.


\begin{table}[ht]
\centering
\begin{footnotesize}
\begin{tabular}{|l|l|l|l|}

\hline

condenser & seed $d$ & overhead  $\delta $& entropy loss $e$ \\
\hline \hline

(1)~\cite[Lemma 4]{bu-mi-ra-ve:c:bitvector} & &     &  \\ non-explicit &  $\log (n/\epsilon) +O(1) $  & $\log (n /\epsilon^2)$ &  $0$ \\ 
\hline
(2)~\cite[Th. 1.7]{guv:j:extractor}& $(1+1/\alpha)(\log (n/\epsilon) + \log k_{max})+O(1)$ &  & \\ 
explicit, strong, linear & any constant $\alpha > 0$ &$\alpha \cdot k_{max}+ 2d$ & $0$\\ 
\hline
(3)~\cite[Th. 2.11]{bau-zim:t:univcodes}\tablefootnote{The condenser (3) appears  inside the proof  of~\cite[Th. 2.11]{bau-zim:t:univcodes} and is a composition of condensers from~\cite[Th. 3.2, also Th.4.1]{tas-uma:c:extractor} and~\cite[Th. 4.3, also Th. 1.7]{guv:j:extractor}. \cite{bau-zim:t:univcodes} is a work in progress, and in the final version this condenser will appear in an explicit form.} & &  &\\ 
    
 explicit, strong, linear & $O(\log(n/\epsilon))$  &  $O(\frac{k_{max}}{\log n} \cdot \log(1/\epsilon) + \log (n/\epsilon))$  &   $0$ \\
\hline

(4)~\cite[Prop. 2.8]{bau-zim:t:univcompression} &  & & \\ explicit, strong & $O(\log k_{max} \cdot \log(n/\epsilon))$ & $0$ &  $d$\\
\hline

(5)~\cite[Th. 22(2)]{rareva:j:extractor} & &  & \\ explicit, strong, linear & $O(\log k_{max} \cdot \log^2(n/\epsilon))$   & $0$ &  $O(d)$ \\

\hline
(6)~\cite{cap-rei-vad-wig:c:conductors}\tablefootnote{Also see~\cite[Th. 10.4]{hoo-lin-wig:j:expander}.} & & & \\ explicit & $O(n - k_{max} + \log (1/\epsilon) - O(1))$ & $d+ \log (1/\epsilon) + O(1)$ & $0$ \\
\hline
\end{tabular}
\caption{\small{The notation represents functions $C: \zo^n \times \zo^d \mapping \zo^{k_{max}+\delta}$, that are  $k \rightarrow_{\epsilon} k+d-e$ condensers for all $k \le k_{max}$ such that $2^k \in \nat$; $n$ is an arbitrary positive integer, $\epsilon > 0$  and $k_{max}$ is any positive integer  $\leq n$.  The corresponding graph $G = (L \cup R, \E)$, where $L = \zo^n, R = \zo^{k_{max}+ \delta}$, with left degree $D:=2^d$, admits $((1-4\epsilon) D, 2 \cdot k_{max} \cdot 2^e)$ online matching up to size $K_{max} := 2^{k_{max}}$.}}
\label{t:tablecondenser}
\end{footnotesize}

\end{table}

\begin{table}[ht]
\centering
\begin{footnotesize}
\begin{tabular}{|l|l|l|}

\hline

bipartite expander & left degree $D'$ &size of the right side $|R'|$   \\
with online matching &  &  \\
\hline \hline
(1)  & & \\ non-explicit &  $O(n^2 k_{max} (1/\epsilon)^2)$ &  $K_{max} \cdot \poly(n/\epsilon)$    \\ 
\hline
& &  \\
(2) & $O( (n/\epsilon)^{4+ (2/\alpha)})$, & $K_{max}^{1+\alpha} \cdot  O((n/\epsilon)^{8+ (4/\alpha)})$    \\ 
explicit, linear & any constant $\alpha > 0$ &  \\ 
\hline
(3)  & & \\ explicit, linear & $\poly(n/\epsilon)$  & $K_{max}^{1+O (\log (1/\epsilon) /\log n)} \cdot \poly(n/\epsilon)$  \\ 
\hline

(4) & & \\ explicit & $2^{O( \log (n/\epsilon) \cdot \log k_{max} )}$ & $K_{max} \cdot  2^{O( \log (n/\epsilon) \cdot \log  k_{max})}$ \\ 
\hline

(5)  & & \\  explicit, linear & $2^{O( \log^2 (n/\epsilon) \cdot \log k_{max} )}$ & $K_{max} \cdot  2^{O( \log^2 (n/\epsilon) \cdot \log  k_{max})}$ \\ 
\hline

(6) & & \\ explicit & $2^{O(n - k_{max} + \log (n /\epsilon))}$ & $K_{max} \cdot   2^{O(n - k_{max} + \log (n/\epsilon))}$ \\
\hline
\end{tabular}
  \caption{\small{The notation represents bipartite graphs $G' = (L' \cup R', \E')$, where $L' =\zo^n$, with left degree $D'$,  that admit $((1-\epsilon) D', 1)$ online matching up to size $K_{max}:=2^{k_{max}}$; $n$ is an arbitrary positive integer, $\epsilon > 0$  and $k_{max}$ is any number such that $K_{max} \in \nat$.  The graphs are obtained from the corresponding graphs $G$  in Table~\ref{t:tablecondenser} via the transformation $G \mapsto G'$ from Section~\ref{s:nosharing}.}}
\label{t:exponlinematching}
\end{footnotesize}

\end{table}


\section{Applications of lossless expanders with online matching}\label{s:applications}
\subsection{One-probe storage scheme for dynamic sets}\label{s:storage}
Recall from the Introduction that the goal is to  store a  subset $S$ of size $K$ of a large set $U$ (the ``universe") of size $N$ and that  
a \emph{one-probe storage scheme} is a data structure  that answers any membership query ``Is $x$ in $S$?" by reading a single bit. 
Buhrman, Miltersen, Radhakrishnan, and Venkatesh~\cite{bu-mi-ra-ve:c:bitvector} have used  lossless expanders to construct randomized one-probe storage schemes  with probability of error bounded by a parameter $\epsilon$.  They have a scheme based on a non-explicit expander that uses a table of size $O(K \cdot \log N \cdot (1/\epsilon)^2)$ bits and an explicit construction achieving table size $O(K^2 \cdot \log N \cdot (1/\epsilon^2))$. Using improved explicit lossless expanders,  there are explicit one-probe storage schemes with smaller tables.  Ta-Shma~\cite{ta-shma:j:storage} obtains table size $K \cdot 2^{O( (\log \frac{\log N}{\epsilon})^3)}$, and Guruswami, Umans, and Vadhan~\cite[Theorem 7.4]{guv:j:extractor} obtain table size $K^\cdot \poly((\log N)/\epsilon) \cdot \exp (\sqrt{\log ((\log N)/\epsilon) \log K})$.  The running times for the explicit schemes are $\poly(\textrm{table size}, \log N)$ for building the table, and $\poly(\log N, \log (1/\epsilon))$ for answering a membership query. 
\if01
\fi

The one-probe storage schemes mentioned above work for representing a \emph{static} set $S$, in the sense that if $S$ changes by inserting or deleting an element, the entire table that represents $S$ has to be recomputed. We show that lossless expanders that admit online matching can be used to represent \emph{dynamic} sets. This means that if an element $x$ is inserted or deleted,  then the table is changed only locally in a few positions without affecting the representation of elements $x'$ different from $x$. This is useful in a distributed environment, because if,  for some $x' \not= x$, the query ``Is $x'$ in S?" is made at the same time  $x$ is inserted or deleted, the table does not have to be locked during the update,  and the data structure still answers the query correctly with probability $1-\epsilon$. The previous sentences give the informal meaning of a one-probe storage scheme for dynamic sets, and, to avoid a tedious discussion, we do not give here a formal definition. But it can  be  inferred from the semantics of the scheme we present below.

\smallskip

We now describe the one-probe storage scheme, presenting its implementation together with its semantics.

For the implementation, we use a lossless expander $G = (L \cup R, \E)$ with left degree $D$, where $L = U$ (the ``universe"), and which admits $((1-\epsilon)D, 1)$ online matching for sets up to size $K+1$.

The idea is simple. The data structure contains a table $T$ of $|R|$ bits, whose entries are indexed by elements in $R$. 
 The bits in $T$ are set so that for every $x \in U$, the bits in the positions assigned to $x$ by the online matching procedure are $1$ if $x \in S$, and $0$ if $x \not\in S$.  Since all but at most an $\epsilon$ fraction of  right neighbors are assigned to $x$, by probing the table at the bit indexed by a random neighbor, we determine if $x$ is in $S$ or not,  with probability $1-\epsilon$. This works because the list of matching requests  has size at most $K$  and  every element that is not in $S$ can be viewed as a dummy element in the last position in the list of matching  requests without going over the allowed bound of $K+1$. So we deduce that the sets of indices assigned to actual elements and virtually to the dummy element  are pairwise disjoint.

We continue with the details. The data structure uses the current \emph{state} of the dynamic set to allow a corect semantics of the operations \emph{insert}, \emph{delete}, and \emph{membership query} in a distributed environment.  The current state is denoted $\tilde{S}$ and it is the \emph{list} of distinct elements from $U$ that have been inserted  during the entire history of the data structure up to the current time, listed in the order in which they were  inserted the first time.   
In addition,  each element in $\tilde{S}$ has a mark which can be either \emph{on} or \emph{off}, indicating if  currently the element is in $S$ or, respectively, has been deleted from $S$.  
The size $|\tilde{S}|$ is the number of distinct elements in the list. For example, if the history of the dynamic set  $S$ is \emph{insert a, insert b, insert c, delete a, delete b, insert a}, then the current state is $\tilde{S} = ((a, \on), (b, \off), (c, \on))$ and its size is $3$. We assume that the total number of inserted elements is bounded by $K$, i.e.,  $|\tilde{S}| \le K$ for all states $\tilde{S}$ during the history of the data structure, and we say that the storage scheme supports up to $K$ insertions. Notice that no element is ever deleted from $\tilde{S}$, and therefore we can update $\tilde{S}$ as a stack, i.e., when some new $x$ is inserted in $S$, then $x$ is pushed in $\tilde{S}$ (and  its mark is set \emph{on};  inserting/deleting elements that have previously been inserted is done by just switching the \emph{on/off} mark). 

The data structure that represents the dynamic set $S$ consists of a table $T$ of bits, having length $|R|$,  whose entries are indexed by the elements of $R$,  and of the state $\tilde{S}$, which can be represented with $O(K \log N)$ bits. Thus, the total size is $|R| + O(K \log N)$.

We use $\tilde{S} + x$ to denote the list obtained by appending $x$ at the end of $\tilde{S}$ if $x \not \in \tilde{S}$, and the list $\tilde{S}$ itself in case $x \in \tilde{S}$. Let $f$ be an assignment function satisfying the requirements in Definition~\ref{d:onlinematching} for $G$.  
To keep the notation simple, $f$ views a state $\tilde{S}$  as a list of elements from $U$ (i.e., $f$ ignores the \emph{on/off} marks). 
Then for every state $\tilde{S}$ of size at most $K$, for every $z \not \in \tilde{S}$ and for every pair $u,v$ of different elements in $\tilde{S}+z$,
\begin{enumerate}
\item[(i)] $f(\tilde{S}+z, u) \subseteq N(u)$,
\item[(ii)]  $|f(\tilde{S}+z, u)|  \ge (1-\epsilon) \cdot D$,
\item[(iii)]  $f(\tilde{S}+z, u) \cap f(\tilde{S}+z, v)   = \emptyset$,
\item[(iv)]   if $z \not= u$, then $f(\tilde{S}, u) = f(\tilde{S}+z, u)$.
\end{enumerate}
We took into account  that the online matching works up to size $K+1$ and thus $f$ is correct for $\tilde{S}+z$.

We next describe the operations \emph{insert, delete} and \emph{membership query}. Initially, all the entries in $T$ are set to $0$ and the state $\tilde{S}$ is the empty list. 
\smallskip

\newcommand{\ins}{ \quad\quad
\textbf{Insert $x$ in  $S$.} 

Lock $\tilde{S}$.

If $x \not \in \tilde{S}$ push $x$ in $\tilde{S}$.  

Mark  $x$  \emph{on} in $\tilde{S}$.
Compute $f(\tilde{S}, x)$  and set $T(p) \leftarrow 1$ for all $p \in  f(\tilde{S}, x)$.

Release $\tilde{S}$.

}

\mybox{gray}{\ins}
\smallskip

 If the lossless expander is explicit, and $f$ is as in Theorem~\ref{r:lexpamatching}, then the running time for \emph{insert} is $\poly(K, D, \log N, 1/\epsilon )$.\footnote{Bruno Bauwens [private communication, Feb 2021] has shown  that it is possible to implement insert and delete in $\poly(\log K, D, \log N, 1/\epsilon)$ amortized time.}
\smallskip

\smallskip
\newcommand{\delx}{\quad\quad \textbf{Delete  $x$ from $S$.} \\
Lock $\tilde{S}$.\\
If $x \not \in \tilde{S}$, do not do anything. \\
Else: Mark $x$ \emph{off} in $\tilde{S}$.  
 Compute $f(\tilde{S}, x)$  and set $T(p) \leftarrow 0$ for all $p \in  f(\tilde{S}, x)$.\\
Release $\tilde{S}.$}
\mybox{gray}{\delx}

\smallskip

 If the lossless expander is explicit, and $f$ is as in Theorem~\ref{r:lexpamatching}, then the running time for \emph{delete} is $\poly(K, D, \log N, 1/\epsilon)$.
\medskip

\mybox{gray}{\quad\quad  \textbf{Membership  query ``Is $x$ in $S$?"} \\
Pick $p$ uniformly at random in $\N(x)$. \\ If $T(p)=1$, answer ``yes"  otherwise answer ``no."}
\smallskip

If the lossless expander $G$ is explicit, the running time for a membership query  is $\poly(\log N, \log D)$ and only one bit of  $T$ is read.
\medskip

We now discuss the semantics of the three operations. Note that during the execution of \emph{insert} and \emph{delete} the state $\tilde{S}$ is locked. This implies that the update operations execute sequentially which guarantees that for every $x \in U$ and at every moment in the history of the data structure, 
\begin{equation}\label{e:sem}
x \in S \Leftrightarrow (x \in \tilde{S} \text{ and its mark is $\on$}).
\end{equation}
 On the other hand, the \emph{membership query} operation does not use $\tilde{S}$ and thus can be executed at the same time with an update operation.  The flip side is that the state $\tilde{S}$ can change during the execution of the \emph{membership query} operation.

We show that the \emph{membership query}  has the following semantics.
\begin{claim} For every $x \in U$, if during the execution of the  \emph{membership query} ``Is $x$ in $S$?" no operation \emph{insert/delete} $x$ is executed and the size of $\tilde{S}$ remains at most $K$, then the answer is correct with probability $1-\epsilon$.
\end{claim}
\begin{proof}
 The assumptions of the claim imply that the locations in $T$ assigned to $x$ are not touched during the execution, and therefore for every state $\tilde{S}$ during the execution of the membership query:
\begin{enumerate}
\item[(a)] If $x$ is in $\tilde{S}$ and its mark is $\on$, then  $T(p) = 1, \text{ for all }  p \in f(\tilde{S}, x) =  f(\tilde{S}+x, x)$,
\item[(b)] If  $x$ is in  $\tilde{S}$ and its mark is $\off$, then $T(p) = 0, \text{ for all }  p \in f(\tilde{S}, x) =  f(\tilde{S}+x, x)$,
\item [(c)] If $x$ is  not in $\tilde{S}$, then $T(p) = 0, \text{ for all }  p \in f(\tilde{S}+x, x)$. 
\end{enumerate}

Let $p$ be chosen at random in $\N(x)$ in the graph $G$. Let ${\cal A}$ be the event  ``$p$  is in $f(\tilde{S}+x, x)$."  Properties (a), (b), (c) and relation~\ref{e:sem}  imply that for every $\tilde{S}$ during the execution, conditioned on the event ${\cal A}$,  
\[
 T(p) = 1 \Leftrightarrow  x \in \tilde{S} \text{ and its mark is on}  \Leftrightarrow  x \in S 
\] 
The conclusion follows because the event ${\cal A}$ has probability at least $1-\epsilon$ (by properties $(i),(ii)$)) for every $\tilde{S}$ that is current during the execution.  

\end{proof}

\smallskip

If we use the lossless expanders from Table~\ref{t:exponlinematching}, we obtain one-probe storage schemes for dynamic sets  with the above semantics,  with various parameters. In particular, we obtain the storage scheme in the next theorem.
\begin{theorem}[{\color{blue}{ one-probe storage scheme for dynamic sets }}  ] \label{t:storedynamic}
For every  functions $k:= k(n) \le n$, $\epsilon:= \epsilon(n) > 0$,  there exists a one-probe storage scheme for dynamic subsets of $U = \zo^n$, supporting up to $K:=2^{k}$ insertions, with table 
size~~$K \cdot 2^{O(\log (n/\epsilon) \cdot \log k)}$. 

The running times are
$\poly(K,  2^{O(\log (n/\epsilon) \cdot \log k)}, 1/\epsilon)$ for the insert/delete operations, and $\poly(n, \log 1/\epsilon)$ for  membership query.
\end{theorem}
\begin{proof}
These parameters are obtained if the above one-probe storage scheme is implemented using the lossless expander (4) from Table~\ref{t:exponlinematching}.
\end{proof}
Except for the case of relatively  small $K$, the parameters in Theorem~\ref{t:storedynamic}  are  as good or better  than in the storage schemes in~\cite{bu-mi-ra-ve:c:bitvector,ta-shma:j:storage, guv:j:extractor}, which  have the limitation of only handling static sets. \footnote{The table size obtained in~\cite{guv:j:extractor} is smaller than the table size in Theorem~\ref{t:storedynamic} in case $\log K/ (\log \log K )^2 < c \log ((\log N)/\epsilon)$ for some constant $c$. The size of the explicit table in~\cite{bu-mi-ra-ve:c:bitvector} is smaller than the table size in Theorem~\ref{t:storedynamic} in case $\log K/ \log \log K < c' \log ((\log N)/\epsilon)$ for some constant $c'$.} 
In this assessment,  we are comparing the table sizes, the running time for  insert/delete in Theorem~\ref{t:storedynamic} vs. the running time for bulding the table for the static case in the earlier schemes, and the running time for the membership query.  
\medskip

\textbf{One-probe storage scheme for a stack.} In case the dynamic set $S$ behaves like a stack (i.e., the element that is deleted is the last element that has been inserted), then a lossless expander yields  a one-probe storage  scheme for $S$ which  works provided  $S$ has at most $K$ elements at each time during its history (note that the number of total insertions can be larger than $K$). This  works in the same way as the scheme above, except that  we use $S$ itself in the role of $\tilde{S}$. Here we use the fact that the assignment function $f(S,x)$ satisfying Definition~\ref{d:onlinematching} allows $S$ to be a stack. 
\medskip


\subsection{Nonblocking networks  with online routing}
 The first application of expanders has been to design networks that have some type of low congestion~\cite{bas-pin:j:networks,mar:j:concentrators}. Unsurprisingly, expanders that admit online matching can be used to design such networks in which routing requests can be satisfied in an online manner.

To illustrate, we consider the following problem. The goal is to construct a directed graph $G$, in which we distinguish  a set $V_1$ of  $N$  
vertices called \emph{input terminals}, and  a set $V_2$  of $N$ vertices called \emph{output terminals}, $V_1 \cap V_2 = \emptyset$,  such that for any sequence  (the ``routing requests") of  $K$  pairs  $(u_1, v_1)$, $(u_2, v_2)$, $\ldots, (u_K, v_K)$ in $V_1 \times V_2$, with all $u_i$'s distinct, and all $v_i$'s distinct, there exist $K$  vertex-disjoint paths from $u_i$ to $v_i$ for all $i \in [K]$.  Such graphs are called \emph{rearrangeable networks}, and we refer to $K$ as \emph{bandwidth}, and to the length of the longest path between input and output terminals  as \emph{depth}.  If the graph can handle online routing, then it is called a \emph{non-blocking network}. Online routing means that the $K$ pairs arrive in order, and when the $i$-th pair arrives,  the path from $u_i$ to $v_i$ has to be established before seeing the future arrivals, i.e. the path from $u_i$ to $v_i$ only depends on the paths from $u_j$ to $v_j$, for $j  < i$.

A simple solution is to take $G$ to  be the  complete bipartite graph with left side $V_1$ and right side $V_2$. The bandwidth is $N$ and the depth is $1$, but the number of edges is $N^2$.  Bassalygo and Pinsker~\cite{bas-pin:j:networks} showed via the probabilistic method the existence  of non-blocking networks with bandwidth $N$, depth $O(\log N)$, $O(N \log N)$ edges, and constant degree. Margulis~\cite{mar:j:concentrators} gives an explicit construction, and there have been many papers studying various variants of the problem (see~\cite{aro-lei-mag:j:network} and the references therein).

\begin{figure}[h]
 \centering{
\begin{tikzpicture}[shorten >=1pt,scale=0.15,
lnode/.style={fill=black, circle, scale=0.2} 
]

\coordinate (v1) at (0,0);
\coordinate (w1) at (12,0);
\coordinate (w2) at (30,0);
\coordinate (v2) at (42,0);

\coordinate (p1) at (0,-2);
\coordinate (p2) at (12,1);
\coordinate (p3) at (30,-3);
\coordinate (p4) at (42,0);

\node[lnode] (pp1) at  (p1){};
\node[lnode] (pp2) at  (p2){};
\node[lnode] (pp3) at  (p1){};
\node[lnode] (pp4) at  (p1){};

\filldraw[opacity=0.6, color=black, fill = gray!70, thick ] (v1) ellipse (5 and 12);
\filldraw[opacity=0.6, color=black, fill = gray!70, thick ] (w1) ellipse (4 and 9);
\filldraw[opacity=0.4, color=black, fill=gray!30, thick, rounded corners] (-7,-13) rectangle (17,14);
\filldraw[opacity=0.6, color=black, fill = gray!70, thick ] (w2) ellipse (4 and 9);
\filldraw[opacity=0.6, color=black, fill = gray!70, thick ] (v2) ellipse (5 and 12);
\filldraw[opacity=0.4, color=black, fill = gray!30, thick, rounded corners] (25,-13) rectangle (49,14);
\draw [dashed, thick, rounded corners](7,-9.5) rectangle (35, 9.5);
\draw[->, thick] (p1)--(p2);
\draw[->, thick] (p2)--(p3);
\draw[->, thick] (p3)--(p4);
\node[anchor=north] (u1) at (p1.-90) {{$u_i$}};
\node[anchor=north] (ww1) at (p2.-90) {$w_i$};
\node[anchor=north] (ww1) at (p3.-90) {{$w'_i$}};
\node[anchor=north] (vv1) at (p4.-90) {{$v_i$}};

\node (cc1) at (0,9) {$\mathbf{V_1}$};
\node (cc2) at (12,6.5) {$\mathbf{W_1}$};
\node (cc1) at (30,6.5) {$\mathbf{W_2}$};
\node (cc1) at (42,9) {$\mathbf{V_2}$};

\end{tikzpicture}
}
\caption{Non-blocking network of depth $3$. The gray rectangles are lossless expanders admitting online matching, and the dashed rectangle is a complete bipartite graph.}
\label{fig-network}
\end{figure}
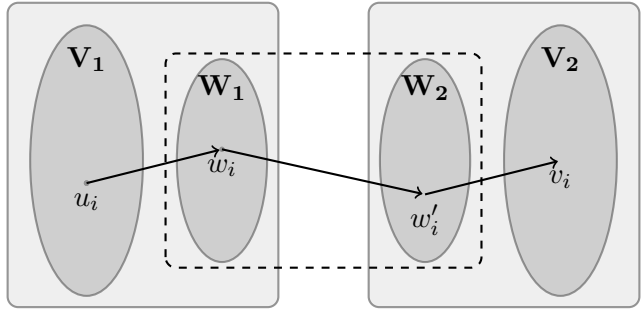

Lossless expanders that admit online matching yield the following simple construction (see Figure~\ref{fig-network}). Take two copies $G_1 = (V_1 \cup W_1, \E_1), G_2 = (V_2 \cup W_2, \E_2 )$ of a lossless expander with $V_1$ and $V_2$ having $N$ nodes,  left degree  $D$ and admitting $((1-\epsilon) D, 1)$ online matching up to size $K$ (the gray rectangles in the figure). Next connect the right sides in all possible ways, i.e., take the complete bipartite graph with sides $W_1$ and $W_2$ (the dashed rectangle). Direct the edges from $V_1$ to $W_1$, $W_1$ to $W_2$, and $W_2$ to $V_2$. We obtain a graph $G$ with $4$ layers: $V_1, W_1, W_2, V_2$, and edges going between layers from left to right. $V_1$ is the set of input terminals, and $V_2$ is the set of output terminals.

 Consider a sequence of $K$ routing requests $(u_1, v_1), \ldots (u_K, v_K)$.   The $i$-th request $(u_i, v_i)$ is satisfied by the following path:   
\[
u_i \longrightarrow \textrm{a node in $W_1$  assigned to $u_i$} \longrightarrow  \textrm{a node in $W_2$  assigned to $v_i$} \longrightarrow v_i. 
\]
More formally, we take the lists $S_1 = (u_1, \ldots, u_K)$ and $S_2 = (v_1, \ldots, v_K)$, 
next we take $w_i$ in $f(S_1, u_i)$ and $w'_i$ in $f(S_2, v_i)$ (where $f$ is the assignment function satisfying the  online matching requirements from  Definition~\ref{d:onlinematching}), and the path is $u_i \longrightarrow w_i \longrightarrow w'_i \longrightarrow v_i$. Clearly, the routing consists of $K$ vertex-disjoint paths, and it satisfies the online requirement. Thus,  $G$ is a non-blocking network, with bandwidth $K$ and depth $3$, having $2ND + |R|^2$ edges ($|R|$ is the size of the right side of the lossless expander). For instance, if we use the lossless expander (4) from Table~\ref{t:exponlinematching} and if $K  = O(N^{1/2})$ then the number of edges is bounded by $N \cdot (\log N)^{O((\log \log K))}$ (thus, almost quasilinear in $N$), which, as far as we know, is not achievable  by previous methods for non-blocking networks with constant degree. 

When $K$ is small, routing can also be done probabilistically very fast without running the assignment process: for every $i \in [K]$, we take $w_i$ to be a random neighbor of $u_i$, and $w'_i$ a random neighbor of $v_i$.  With probability $1 - 2K \epsilon$, all the random choices belong to the set of nodes assigned by the online matching, and therefore, the paths are vertex-disjoint. 

 \section{Acknowledgments} The author is grateful to Bruno Bauwens for intense and very helpful discussions.


\bibliography{theory-3}

\bibliographystyle{alpha}

\appendix
\section{Appendix: Connection with Bauwens, Zimand~\cite{bau-zim:t:univcompression}}\label{s:bz19}

We briefly explain how the online matching properties of lossless expanders follow from results in~\cite{bau-zim:t:univcompression}.
\medskip

 ~\cite{bau-zim:t:univcompression} introduces the concept of   $(K,\varepsilon)$-invertible function:

\begin{definition}[Definition 2.1 in~\cite{bau-zim:t:univcompression}] \label{def:invertible}
  A probabilistic function $\ff \colon \mcX \rightarrow \mcY$ is {\em $(K,\varepsilon)$-invertible} if there exists a deterministic partial function 
  \mbox{$g \colon \mcX^{\le K} \times \mcY \rightarrow \mcX$} such that for all $S \in \mcX^{\le K}$ and all~$x \in S$:
 \[
   \Pr \left[g_S(\ff(x)) = x\right] \;\ge\; 1-\varepsilon\,,
 \]
 where $g_S(y) = g(S,y)$.
 $\ff$ is  {\em online} $(K,\varepsilon)$-invertible if there exists such a function $g$ that is {\em monotone} in $S$: 
 if list $S'$ extends $S$,  then the function $y \mapsto g_{S'}(y)$ is an extension of~$y \mapsto g_S(y)$. (Note: $\mcX^{\le K}$ is the set of all sequences of length at most $K$ with elements from the set $\mcX$.)
\end{definition}
If the sets $\mcX$ and $\mcY$ are finite and if the function $\ff$ is using the same amount of randomness, say $d$ bits, for each input in $\mcX$, we can associate to the invertible function in the standard way a bipartite graph, with left degree $D = 2^d$, where the left side is $\mcX$, the right side is $\mcY$, and $(x,p)$ is an edge if there is a random string $\rho$ such that $\ff(x, \rho) = p$. In this view, an invertible $(K, \epsilon)$ function is a bipartite graph that has $((1-\epsilon)D, 1)$ online matching.  In this way results from~~\cite{bau-zim:t:univcompression} about online invertible functions can be translated in the language of this paper to refer to bipartite graphs that admit online matching.

With this translation and taking into account that the conductors defined in ~\cite{bau-zim:t:univcompression} are standard conductors with entropy loss $d$, Corollary 2.13 in~\cite{bau-zim:t:univcompression} states in the language of this paper that a $(K, \epsilon)$ conductor with loss entropy $d$, after a transformation similar to the $G \mapsto G'$ transformation in this paper, produces a graph that admits $(1-O(\epsilon) D, 1)$ online matching up to size $K$.  Therefore a  $(K, \epsilon)$ lossless expander (which is equivalent to a conductor with entropy loss $0$~\cite[Th. 8.1]{ats-uma-zuc:j:expanders}) also    admits $((1-O(\epsilon)) D, 1)$ online matching up to size $K$, and this is  Theorem~\ref{t:main}(b).

In a similar way,  Corollary 2.11 in~\cite{bau-zim:t:univcompression} states that a $(K, \epsilon)$ conductor with loss entropy $d$ corresponds to a graph that admits $((1-2\epsilon) D, D \log (2K))$ online matching up to size $K$. This is almost the same as Theorem~\ref{t:main}(a), except that entropy loss should be $0$ (so that the graph is a lossless expander), and the online matching should have sharing parameter $O(\log K)$ instead of $D \log (2K)$. Some tweaking of the graphs eliminates these differences.

\end{document}